# Peer-to-Peer Multimedia Sharing based on Social Norms

Yu Zhang, Mihaela van der Schaar

Department of Electrical Engineering, UCLA

yuzhang@ucla.edu, mihaela@ee.ucla.edu

*Abstract*—Empirical data shows that in the absence of incentives, a peer participating in a Peer-to-Peer (P2P) network wishes to download content, while avoiding to contribute content in return. This phenomenon, known as free-riding, has been actively studied in the literature and incentives have been proposed to compel self-interested peers to cooperate with each other and contribute their content to other peers in the network. Most solutions for providing incentives in P2P networks are based on direct reciprocity, which are effective in networks where the peers are interested in the same content (e.g. downloading a large file such as the Linux operating system). However, such incentives schemes are not appropriate for most P2P multimedia sharing networks due to the unique features exhibited by such networks: large populations of anonymous agents interacting infrequently, asymmetric interests of peers, network errors, and multiple concurrent transactions. In this paper, to address these challenges, we design and rigorously analyze a new family of incentive protocols that utilizes indirect reciprocity and it is based on the design of efficient social norms. In the proposed P2P protocols, the social norms consist of a social strategy, which represents the rule prescribing to the peers when they should or should not provide content to other peers based on their own reputation as well as the reputation of the requesting peers, and a reputation scheme, which rewards or punishes peers depending on whether they comply or not with the social strategy. We first define the concept of a sustainable social norm, under which no peer has an incentive to deviate. We then formulate the problem of designing optimal social norms, which selects the social norm that maximizes the network performance among all sustainable social norms. We study the structure of optimal social norms and specifically determine for peers of a certain reputation whether they will need to serve peers of a specific reputation and whether, in turn, they would be able to receive service from that reputation or not. In particular, we show that, given the network and peers' characteristics, social norms can be designed which deter free-riders by reducing their reputation and thus, the service which they receive from the P2P network. Hence, we prove that it becomes in the self-interest of peers to contribute their content to the network rather than to free-ride. We also show the importance of designing optimal social norms rather than just selecting a sustainable social norm in an ad-hoc manner. We also investigate the impact of various punishment schemes on the social welfare as well as how should the optimal social norms be designed if altruistic and malicious peers are active in the network. Our results show that optimal social norms are capable of providing significant improvements in the sharing efficiency of multimedia P2P networks.

I. INTRODUCTION

With the explosion of communication technologies and multimedia signal processing, the sharing of multimedia content is becoming increasingly popular over the Internet. In particular, Peer-to-Peer (P2P) multimedia applications represent a large majority of the traffic currently exchanged over the Internet. By pooling together the resources of many autonomous devices, P2P networks are able to provide a scalable and low-cost platform for disseminating large files without relying on a centralized infrastructure [1][2]. Multimedia sharing systems that have been successfully developed for P2P networks usually use tree-based or data-driven approaches [4][5]. In this paper, data-driven approaches adopting pull-based techniques are considered [4][6], where different types of files are divided into chunks of uniform length and are then disseminated over the P2P network. Each peer possesses several chunks, which are shared among interested peers, and information about the availability of the chunks is periodically exchanged among peers through intermediate trackers. Using this information, peers continuously associate themselves with other peers and exchange chunks.

While this approach has been successfully deployed in various applications over P2P networks, it is vulnerable to intrinsic incentive problems since the upload service incurs costs to both the uploader and the downloader, but benefits only the downloader [6]. As contributing their content does not generate direct benefit, peers tend to avoid uploading while trying to download content from other peers, a behavior commonly known as *free-riding*.

Such studies demonstrate that designing incentive protocols to encourage cooperation and mitigate free-riding is crucial to maintain the performance of P2P multimedia sharing applications. To achieve this goal, a large body of research was dedicated to this area [7][8]. Many of these existing mechanisms rely on game-theoretical approaches and can be classified into two categories: pricing and reciprocity [9].

Pricing mechanisms rely on implementing a currency-based system that is resistant to forgery and double-spending [10]. Peers are incentivized to share their content by rewarding them with virtual currency for uploading and charging them for downloading. However, such solutions are often very cumbersome to deploy because they require an accounting infrastructure to track the transactions of peers, which further necessitates the usage of public keys, a web of trust, or threshold cryptography techniques [11]. Furthermore, these systems often deploy auctions to set the price, which may result in high delay and complexity in order to implement a desirable allocation.

Another method for providing incentives is based on reciprocity, where the peers' past reciprocative behavior (e.g. contributing content to other peers or not) is rewarded or punished in future interactions with the same or other peers. Differential service schemes are deployed in reciprocity-based protocols to determine how peers should make their upload decisions: a peer with a higher rating (i.e. a peer that exhibited a good past behavior) receives more resources than a peer with a lower rating [12]. Since a peer with a high rating is treated preferentially by other peers, incentives are provided for peers to cooperate in order to build up high ratings. Depending on how a peer's rating is generated, reciprocity-based protocols can be classified as direct reciprocity (also known as personal reciprocation) and indirect reciprocity (also referred to as societal reciprocation).

In direct reciprocity, each peer rates a specific peer individually [6]. Hence the interaction between two peers is only influenced by their own history of interactions. Though easy to implement, direct reciprocity requires frequent interactions between two peers in order to establish accurate mutual ratings, which is restrictive in P2P networks characterized by high churn or asymmetry of interests. For example, the investigation in [2] shows that over 70% of P2P traffic is exchanged in networks with more than 1000 peers, which implies that a peer normally interacts with a stranger (i.e. with whom it was randomly matched) about whom it has no prior history and with whom it has no expectation to meet again in the future. Hence, protocols based on direct reciprocity such as tit-for-tat perform well only in networks dominated by long-lived relationships, where peers have ample opportunities to mutually reciprocate, and where peers are interested in similar content.

Due to the random matching feature of large P2P networks, indirect reciprocity becomes a more appropriate mechanism in designing incentive protocols. Most protocols based on indirect reciprocity use reputation mechanisms [14][15][16]. A peer is globally rated with a reputation calculated by its past behavior in the network. In order to make a decision, a peer does not need to know the entire action history but the reputation of its opponent. However, the majority of existing works on P2P reputation mechanisms are concerned with system design issues and focus on effective information gathering techniques which only differ in how the global reputation is calculated and propagated, e.g. efficient information aggregation [15], secure peer identification [16], etc. An analytical framework that is able to rigorously study how peers can be incentivized to cooperate in P2P networks and what is the resulting impact on the network performance when various reputation mechanisms are deployed, is still missing.

In our past work [22], we developed a rigorous framework for studying reputation mechanisms which can be applied to P2P applications. The peers determine their upload services to a specific peer based on this peers' reputation, as well as their own status in the P2P system (i.e. their own reputations). To formalize the reputation mechanism, *social norms* [17], which consist of a social strategy and a reputation scheme, are introduced to regulate the behavior of peers. We derived analytically under what conditions peers will find in their own self-interest to comply with the prescribed social strategy, and defined and solved the protocol designer's problem of designing an optimal social norm that maximizes the social welfare of the network.

In this paper, we use the theoretical framework developed in [22] to design efficient incentive protocols for P2P multimedia sharing services. Hence, we design social norms, which consider the unique features and constraints of P2P multimedia sharing services:

- *Asymmetry of interests among peers*. In multimedia sharing applications, peers are interested in very diverse contents. Hence, a peer providing content to a requesting peer may not be interested in any of the content possessed by this peer, but it may be interested in content available from other peers. To accommodate the fact that the peers' interests are asymmetric, we model the interaction between a pair of matched peers as a gift-giving game, instead of a prisoner's dilemma game, which assumes mutual interests between a pair of peers and is widely adopted in traditional analysis of P2P systems.

- *Service errors*. The existing literature on incentives for P2P networks [14][15] rarely considers that network errors may affect the interactions between peers. This is an idealized assumption which is hard to realize in practical networks. In contrast, our work explicitly takes into consideration that the exchange of chunks between peers may be subject to service errors and considers how protocols can be efficiently designed given the level of network errors.

- *Multiple connections*. In multimedia sharing applications over P2P networks, peers can engage in multiple simultaneous connections with other peers to exchange chunks in order to increase the download efficiency. Hence, we augment the framework in [22] to accommodate sharing using multiple connections and explicitly analyze how the number of connections will impact the peers' incentives and the social welfare of the P2P network.

- *Simple protocol designs*. Unlike in [22], where our focus is on determining structural results for the most efficient social norms irrespective of their resulting implementation and designing complexity, in this paper we restrict our attention to a simpler class of social strategies – the set of threshold-based strategies, in which peers can receive services only if their reputations are higher than a threshold.

- *Altruistic peers and malicious peers*. We also rigorously determine the impact of altruistic peers (who always provide upload services to other peers) as well as malicious peers (who upload corrupted data to other peers) on the protocol design and the P2P network performance.

The remainder of the paper is organized as follows. In Section II, a game-theoretic model for P2P multimedia sharing is proposed. In Section III, the problem of designing the optimal incentive protocol is formalized and the structure of the optimal protocol is studied. Section IV explicitly investigates the impact of altruistic and malicious peers on the performance and robustness of the incentive protocols. After showing the simulation results with illustrative examples in Section V and discussing the possible future extensions of our work in Section VI, we conclude the paper in Section VII.

## II. SYSTEM MODEL

*A. Considered P2P networks*

We consider a P2P multimedia sharing network such as Chainsaw [3] and CoolStreaming [4] [1], where peers would like to associate themselves with other peers that possess media content in which they are interested. The shared media content is coded and divided into media chunks by the content creator. Here we define the value (benefit) of a chunk as its dependency factor on other chunks, which represents the video distortion reduction on the peer who receives this chunk [25]. In general, both the value and the size of a chunk may depend on the priority class to which it belongs to. To make the analysis tractable, we assume initially that the multimedia content is encoded such that all the chunks are of equal size and have the same value (benefit). In Section VI, we will discuss the case when chunks are classified into different priority classes, which are defined based on the specific video encoder used by the content creator (e.g. they could be base and enhancement layers like in a scalable video coder; the I, P and B-frames of an H.264/AVC codec etc.).

---

[1] The results obtained in this paper can be applied in P2P applications other than multimedia sharing, such as overlay routing and general file sharing without change.

At any instance, a peer buffers an amount of chunks that can be shared with others, and the trackers maintain and update periodically the buffer maps recording the content possession of each peer. We consider a discrete-time model, in which time is divided into periods representing the interval between two updates of the buffer maps by the trackers.

We assume that there is a continuum of peers in the network, which is a good practical model for large-scale P2P networks [1][14]. When a peer wants to download a certain chunk, it sends a search request to the tracker from which it receives a response with the list of peers who have the requested content [6]. Then the peer randomly selects a peer from the list to send a service request. The selection is uniformly random such that all peers on the list have an equal probability to be chosen [18][20]. At any instance, an individual peer can support simultaneously a fixed number of $b$ download connections, from which it downloads chunks it requested from others [6].

*B. The stage game played by a pair of peers*

The interaction between a pair of connected peers exchanging a chunk, which is defined as a *transaction*, can be modelled as a one-stage asymmetric gift-giving game to characterize the asymmetry of interests among peers [21]. To avoid confusion, the peer who requests the downloading of a chunk is called a *client* and the peer who is being requested is called a *server*. In one transaction, the server has the choice of selecting its action $a$ from the set $\mathcal{A} = \{S, NS\}$, where $S$ stands for "Serve", implying that the server responses to the client's request to upload the chunk; whereas $NS$ stands for "Not Serve", implying that the server refuses to upload the chunk. The utility matrix of one transaction is illustrated in Table 1 which is specified as follows.

- If $a = S$, the server consumes an upload cost of $c$, and the client receives a benefit of $r$. $c$ can be determined as a composite function of the upload bandwidth and the transmission power spent by the server [18]. It should be noted that this cost may be different for various peers, and can vary over time and per chunk. In our formal analysis, we consider that $c$ is constant for each chunk, but our proposed framework can be extended to take peer-dependent and time-varying costs into consideration. The value (benefit) of the chunk $r$ represents the video distortion reduction that the client experiences from receiving a specific chunk. This may also vary depending on its content, the priority class to which it belongs to etc. As for the cost, we consider $r$ to be fixed in our analysis, but extensions to variable benefits can be developed within the framework proposed here. We assume that $r > c$ such that the sharing service provided by the P2P network is socially valuable.

- If $a = NS$, both the server and client receive a utility of 0.

Since each peer can maintain multiple simultaneous connections, the utility it receives in one period is the sum utility from all the transactions in which it is involved or, in other words, from all of its established upload and download connections. The social welfare of the network is quantified by the social utility $U$ that is defined as the average utility of all peers in one period. We assume that peers in the network are self-interested and aim to maximize their individual utilities and therefore, they will only upload chunks if this has a positive impact on their future utilities (e.g. they can increase their future downloads). Since $r > c$, the social utility is maximized when all servers choose $a = S$ in their transactions. Nevertheless, the myopic equilibrium of the one-stage gift-giving game is $a = NS$, with which a self-interested server who has the incentive to free-ride can maximize its stage-game utility myopically, which gives rise to an undesirable utility of 0 for each transaction and hence a zero social utility for the entire network.

*C. Social norms*

We adopt a repeated game formulation to model the subsequent interactions among peers and we adopt P2P protocols based on social norms in order to improve the inefficiency of the myopic equilibrium. Social norms define the rules that the group of peers uses to reward or punish appropriate and inappropriate behaviors in the P2P network. Since we focus on protocols that are based on social norms, we use the two terms "protocol" and "social norm" interchangeably in the rest of the paper.

In the repeated game, each peer is tagged with a reputation $\theta$ representing its social status. We consider the case where reputations take values from a finite set. Thus, without loss of generality, we assume that $\theta$ is a natural number from the finite set $\Theta = \{0, 1, 2, \cdots, L\}$ for some $L$. For notational convenience, a peer of reputation $\theta$ is referred to as a $\theta$-peer. The high reputation of a peer reflects its cooperative behavior in the past, i.e. this peer uploaded content to peers requesting it. The highest reputation $L$ can be gained by a peer which has been cooperative in the past $L$ periods. The reputation of the peers is maintained and updated by a trustworthy third-party device, e.g. the tracker.

The social norm, denoted by $\kappa$, is determined by the P2P protocol designer, which is composed of a social strategy $\sigma$ and a reputation scheme $\tau$.

$\sigma$ is a *reputation-based behavioral strategy*, which is represented by a mapping $\sigma : \Theta \times \Theta \to \mathcal{A}$, where the first $\Theta$ represents the server's reputation, the second $\Theta$ represents the client's reputation, and $\mathcal{A}$ represents the server's action. It specifies what action $\sigma(\theta, \tilde{\theta}) \in \mathcal{A}$ should a server of reputation $\theta \in \Theta$ select when meeting a client of reputation $\tilde{\theta} \in \Theta$.

$\tau$ serves as the reward and punishment system in the social norm, and it specifies how a peer's reputation should be updated based on its actions in the transactions that it is engaged. In our framework, $\tau$ updates a peer's reputation at the beginning of each period. Specifically, the tracker reviews all upload transactions of a peer with the result of the review recorded in a variable $x \in \{0,1\}$. At the beginning of a period, $x$ is reset to 0. Then in each transaction, there is a mapping $\phi$ which maps the reputations of the serving peer and its client as well as the serving peer's action during one transaction into a binary value as $\phi: \Theta \times \Theta \times \mathcal{A} \to \{0,1\}$. If the action is in accordance to the social strategy, $\phi$ outputs 0 indicating that the peer behaves well in this transaction; otherwise, if the action is against the social strategy, $\phi$ outputs 1 indicating that the peer does not behave well by not complying with the social norm. After the transaction, $x$ is updated by an OR-operation as $x := x \vee \phi$. That is, the new value of $x$ will be 0 if and only if both $\phi$ and the old value of $x$ is 0. Hence, after one period, $x = 0$ if and only if the peer complies with the social norm in all of its upload transactions. Once it has deviated in any transaction, we have $x = 1$. Based on the peer's current reputation and $x$, $\tau$ then determines its new reputation as $\tau: \Theta \times \{0,1\} \to \Theta$. The mapping rule is as follows: if $x = 0$, $\tau$ rewards the good behavior of the peer with an increased reputation; on the other hand, if $x = 1$, $\tau$ punishes the peer for not uploading sufficient content in this period with a decreased reputation.

In our framework, a peer's upload action in one transaction that is used in the mapping of $\phi$ is reported by its client. We here assume that the client always makes a truthful report [2]. However, we do consider the impact of network (service) errors. With the probability $\varepsilon$ ($0 < \varepsilon \ll 1$), a peer which intends to upload a chunk in one transaction fails to do so due to a connectivity error.

To encourage cooperation among peers, we restrict our attention to a set of threshold-based strategies $\Gamma$. Every strategy $\sigma \in \Gamma$ can be characterized by a service thresholds $h(\sigma) \in \{1, \cdots, L\}$ [3], which can be specified as follows (We relax this constraint that one strategy only has one service threshold in Section III, where difference peers have different service thresholds.)

$$\sigma(\theta, \tilde{\theta}) = \begin{cases} S & \text{if } \theta \geq h(\sigma) \text{ and } \tilde{\theta} \geq h(\sigma) \\ NS & \text{otherwise} \end{cases}, \quad (1)$$

By adopting $\sigma$, peers with reputation being at least $h(\sigma)$, which are called "active peers", will mutually help each other, while peers with reputation lower than $h(\sigma)$, referred to as "inactive peers", cannot download chunks from others and do not need to upload chunks to others. To avoid confusion, the prescribed social strategy is denoted as $\sigma_o$ and the corresponding prescribed service threshold of the social strategy is denoted as $h_o$ in order to differentiate these from an ordinary behavioral strategy.

---
[2] The extension to the untruthful report from clients is discussed in Section VI.
[3] Here the strategies with the service threshold being 0 and *L+1* are not considered.

To encourage compliance with the social strategy which will induce cooperative behaviors among peers, the peers are rewarded by an increased reputation. Incompliance (i.e. not complying with the prescribed social strategy) is punished by a decreased reputation. To keep the initial design of the P2P protocol simple, we consider a reputation scheme $\tau$ that provides the harshest punishments to peers when they do not comply with the social strategy. (We relax this constraint in Section III, where incompliance to the social norm does not necessarily lead to a peer's reputation falling to 0.) The reputation update rule can be written as follows

$$\tau(\theta, x) = \begin{cases} \min\{L, \theta+1\} & if\ x = 0 \\ 0 & if\ x = 1 \end{cases}. \tag{2}$$

A schematic representation of a social norm is provided in Figure 1, with Figure 1 (a) illustrating the decision process of a social strategy, where $\tilde{\theta}$ denotes the reputation of the client in one transaction and Figure 1 (b) illustrating the decision process of a reputation scheme.

Here we briefly explain how a peer's reputation changes after one period under this reputation scheme.
(1) For an inactive peer of reputation $\theta < h_o$, it has to comply with the prescribed social strategy $\sigma_o$ to play $a = NS$ in all of its upload transactions. Since it does not upload chunks to others, there is no error taking place in any of its transactions and hence, its reputation can always be successfully increased by 1 after one period by complying with $\sigma_o$.
(2) The social strategy $\sigma_o$ prescribes that an active peer of reputation $h_o \leq \theta < L$ should play $a = S$ with peers of reputation $\theta \geq h_o$, i.e. other active peers, and it should play $a = NS$ with peers of reputation $\theta < h_o$, i.e. inactive peers. Its reputation will be increased by 1 if it complied with $\sigma_o$ in all the transactions in which it was involved. However, if it failed to upload chunks to an active peer, either deliberately or due to a service error, in any transaction, its reputation falls to 0.
(3) Similarly, if an $L$-peer complies with $\sigma_o$ in all transactions of one period with $x = 0$, this peer will continue to hold reputation $L$.

It should be noted that an inactive peer's reputation is always increased after one period until it reaches $\theta = h_o$ and becomes an active peer. On the contrary, there is always a positive probability of an active peer's reputation not being increased but falling to 0. (We will investigate in a next section what is the impact on less harsh punishments.) A peer's reputation transition probability across periods will be explicitly calculated in the next section.

The sequence of events in one transaction is summarized in Table 2.

*D. Utilities*

As discussed in the above section, an inactive peer receives a utility of 0 in one period. In this section, we determine the utility of an active peer. An active peer's expected one-period utility depends on the rate at which it requests and is requested for chunks. Here we assume that each active peer generates chunk requests at a constant rate [14]. In one period, each peer generates a constant request of $\lambda b$ chunks, where $\lambda$ can be interpreted as the rate at which each connection is utilized per period [13]. Once the download request is rejected, the peer immediately redirects this request to another peer on the list provided by the tracker until it is matched with a peer who accepts its request. Hence, an active peer can always download $\lambda b$ chunks in one period. Due to the random matching feature of the network, the chunks uploaded by an active peer per period is also $\lambda b$ [4]. In summary, the expected one-period utility of a peer can be expressed as

$$v_\kappa(\theta) = \lambda b\left[(1-\varepsilon)r - c\right], \text{ if } \theta \geq h_o, \tag{3}$$

and

$$v_\kappa(\theta) = 0, \text{ if } \theta < h_o. \tag{4}$$

We use the infinite-horizon discounted sum criterion to evaluate a peer's expected overall utility as the sum of its expected one-period utility in the current period and its discounted expected overall utility starting from the next period. Let $p_\kappa(\theta' \mid \theta)$ denote the transition probability of a peer's reputation across periods when following $\kappa$, which can be determined as follows

$$p_\kappa(\theta' \mid \theta) = \begin{cases} 1-\alpha, & \theta \geq h_o \text{ and } \theta' = \min\{L, \theta+1\} \\ \alpha, & \theta \geq h_o \text{ and } \theta' = 0 \\ \alpha, & \theta < h_o \text{ and } \theta' = \theta+1 \\ 0, & otherwise \end{cases}, \tag{5}$$

where $\alpha$ is the probability that an active peer who complies with the social norm is falsely punished due to the service error, i.e. its reputation is decreased to 0 rather than being increased. Since a peer's reputation is increased if and only if it complies with the social norm in all upload transactions within one period, we have $\alpha = 1 - (1-\varepsilon)^{\lambda b}$.

Therefore, a peer's expected overall utility in the repeated game starting from any period $t_0$ when following $\kappa$, can be expressed as

$$v_\kappa^\infty\left(\theta^{t_0}\right) = \mathbb{E}\left[\sum_{t=t_0}^{\infty} \delta^t v_\kappa\left(\theta^t\right)\right] = v_\kappa\left(\theta^{t_0}\right) + \delta\sum_{\theta'} p_\kappa\left(\theta' \mid \theta\right) v_\kappa^\infty\left(\theta'\right), \qquad (6)$$

where $\delta \in [0,1)$ is a peer-defined discount factor, which represents the weight that a peer gives to its utility that can be received in the future.

---

[4] The validity of this conclusion also depends on the assumption that chunks are uniformly distributed among peers in the network and each chunk has the same popularity.

The social utility of the network is regarded the average expected one-period utility over all peers and hence depends on the reputation distribution of the peer population. The reputation distribution in one period is denoted by $\{\eta(\theta)\}_{\theta=0}^{L}$, with each term $\eta(\theta)$ representing the fraction of peers in the total population holding a reputation $\theta$. Due to the reputation update in each period, $\{\eta(\theta)\}$ evolves dynamically over time. Here we assume that $\{\eta(\theta)\}$ is updated at the beginning of each period and is broadcast to all peers in the network. Let $\mu^t$ denote the fraction of active peers in period $t$. It is equivalent to the fraction of peers whose reputations are above $h_o$, and thus can be expressed as $\mu^t = \sum_{\theta=h_s}^{L} \eta^t(\theta)$. Consequently, the update of $\{\eta(\theta)\}$ across periods can be characterized by the following set of equations:

$$\begin{aligned}
\eta^{t+1}(L) &= (1-\alpha)\eta^t(L) + (1-\alpha)\eta^t(L-1) \\
\eta^{t+1}(\theta) &= (1-\alpha)\eta^t(\theta-1), \ h_o+1 \leq \theta \leq L-1 \\
\eta^{t+1}(\theta) &= \eta^t(\theta-1), \ 1 \leq \theta \leq h_o \\
\eta^{t+1}(0) &= \alpha\mu^t \\
\mu^{t+1} &= \sum_{\theta=h_o}^{L} \eta^{t+1}(\theta)
\end{aligned} \qquad (7)$$

Since we are interested in the long-term utilities of peers, we study the distribution of reputations in the long run, which is defined as follows.

*Definition 1 {Stationary distribution}.* $\{\eta_\kappa(\theta)\}$ is a stationary distribution of reputations under the dynamics defined by (7) if it satisfies $\sum_{\theta=0}^{L} \eta_\kappa(\theta) = 1$ and

$$\begin{aligned}
\eta_\kappa(L) &= (1-\alpha)\eta_\kappa(L) + (1-\alpha)\eta_\kappa(L-1) \\
\eta_\kappa(\theta) &= (1-\alpha)\eta_\kappa(\theta-1), \ h_o+1 \le \theta \le L-1 \\
\eta_\kappa(\theta) &= \eta_\kappa(\theta-1), \ 1 \le \theta \le h_o \\
\eta_\kappa(0) &= \alpha\mu_\kappa \\
\mu_\kappa &= \sum_{\theta=h_o}^{L} \eta_\kappa(\theta)
\end{aligned} \qquad (8)$$

It should be noted that each variable in (8) is subscripted with $\kappa$ to highlight its dependence on the particular social norm. As in [22], we prove the existence of and convergence to a unique stationary distribution $\eta_\kappa$ for a particular social norm $\kappa$ when all peers comply with its social strategy $\sigma_o$.

**Lemma 1**. When all peers follow the social strategy $\sigma_o$, the reputation distribution of the network converges to a unique stationary point $\{\eta_\kappa(\theta)\}$ as follow:

$$\begin{aligned}
\eta_\kappa(L) &= 1 - (1+h_o\alpha)\mu_\kappa + (1-\alpha)^{L-h_o}\mu_\kappa \\
\eta_\kappa(\theta) &= (1-\alpha)^{\theta-h_o}\alpha\mu_\kappa, \ h_o+1 \le \theta \le L-1 \\
\eta_\kappa(\theta) &= \alpha\mu_\kappa, \ 0 \le \theta \le h_o \\
\mu_\kappa &= \frac{1}{1+\alpha h_o}
\end{aligned} \qquad (9)$$

*Proof*: See [22]. ∎

Therefore, the social utility of the network is defined as the expected one-period utility averaged over all peers when the reputation distribution is stationary

$$U_\kappa = \sum_\theta \eta_\kappa(\theta) v_\kappa(\theta) = \lambda b \mu_\kappa [(1-\varepsilon)r - c]. \qquad (10)$$

## III. OPTIMAL DESIGN OF SOCIAL-NORM BASED PROTOCOLS

*A. Defining sustainability in P2P networks*

While designing a protocol, the incentive of peers to follow the prescribed social strategy has to be investigated, i.e. we need to investigate whether a social norm is sustainable. Since we consider a non-cooperative scenario, in order to ensure that a peer has no incentive for deviating unilaterally from the social norm, we need to check whether a peer can improve its expected overall utility by deviation. Let $c_\kappa(\theta)$ denote the one-period cost consumed by a server with reputation $\theta$ following the social norm $\kappa$. As the first step, we define a social norm to be the *social norm equilibrium* [17] as follows.

*Definition 2 {Social norm equilibrium}*. The social norm $\kappa = (\sigma_o, \tau)$ constitutes a social norm equilibrium if the sum of its instant utility in one period and its expected future utility thereafter by complying with the social norm $\kappa$ is larger than the sum of utilities by deviating to any other behavioral strategy $\sigma$ [5], i.e.

$$-c_\kappa(\theta) + \delta \sum_{\theta'} p_\kappa(\theta' \mid \theta) v_\kappa^\infty(\theta') \geq -c_\sigma(\theta) + \delta \sum_{\theta'} p(\theta' \mid \theta, \sigma) v_{\kappa,\sigma}^\infty(\theta'), \text{ for all } \theta \text{ and } \sigma \in \Theta, \quad (11)$$

where $c_\sigma(\theta)$, $p(\theta' \mid \theta, \sigma)$, and $v_{\kappa,\sigma}^\infty(\theta)$ are a peer's incurred cost per period, its reputation transition probability from $\theta$ to $\theta'$, and its expected overall utility, respectively when it plays $\sigma$ and the protocol designer implements the social norm $\kappa$ [6].

Hence, if $\kappa$ is the social norm equilibrium, the peer cannot gain from unilateral deviation regardless of the reputation of the client it is matched with when every other peer follows the prescribed social strategy $\sigma_o$. Thus, under a social norm equilibrium, peers will find it in their own self-interest to comply with the social strategy $\sigma_o$.

The definition of social norm equilibrium requires $\sigma_o$'s optimality to be checked using the above definition across all possible behavioral strategies under $\kappa$, thereby incurring a high computational complexity. We establish the one-shot deviation principle for social norm equilibrium in [22] to simplify the computation, which at the same time serves as a sufficient and necessary condition for the existence of the social norm equilibrium.

**Lemma 2 (One-shot Deviation Principle)**. $\kappa$ is a social norm equilibrium if and only if for any $\theta$, there is no profitable one-shot deviation, i.e.

$$c_\sigma(\theta) - c_\kappa(\theta) \leq \delta \left[ \sum_{\theta'} p_\kappa(\theta' \mid \theta) v_\kappa^\infty(\theta') - \sum_{\theta'} p(\theta' \mid \theta, \sigma) v_\kappa^\infty(\theta') \right], \text{ for all } \sigma \in \Theta. \quad (12)$$

*Proof*: See [22]. ∎

Lemma 2 shows that if a peer cannot gain by unilaterally deviating from the prescribed social strategy $\sigma_o$ only in the current period while following $\sigma_o$ afterwards, it also cannot gain by switching to any other strategy $\sigma$, and vice versa. The left-hand side of (12) can be interpreted as the gain that a peer can obtain from the saving on its upload cost in one period by choosing $\sigma$, while the right-hand side of (12) represents the discounted expected future loss which the peer will suffer due to the decrease in reputation incurred by choosing $\sigma$. Using the one-shot deviation principle, we can derive incentive constraints that characterize sustainable social norms. There are two cases that need to be considered.

When an active peer with reputation $\theta$ receives upload requests from another active peer with reputation $\tilde{\theta}$, then the protocol requires the $\theta$-peer to upload the chunk to the $\tilde{\theta}$-peer. Thus, the protocol should provide the $\theta$-peer incentives to choose $a = S$ over $a = NS$. By following the protocol, the $\theta$-peer incurs the upload cost $c$ in this transaction while its reputation in the next period becomes $\min\{L, \theta+1\}$ with a probability that is at least $(1-\alpha)$ and $0$ with a probability that is at most $\alpha$. Meanwhile, it expects to receive $M$ more upload requests other active peers with $M \leq \lambda b - 1$. Thus, the resulting expected overall utility of the $\theta$-peer is given by $V_\theta(S) = -(1+M)c + \delta\left[(1-\alpha)v_\kappa^\infty(\min\{L, \theta+1\}) + \alpha v_\kappa^\infty(0)\right]$. On the contrary, if the peer deviates by refusing to upload the requested chunk and play $a = NS$, it saves an instant cost of $c$ in this transaction as well as in all the future transactions within this period [7], but at the expense that its reputation falls to $0$ with probability 1, starting from the next period [8]. The expected overall utility is thus $V_\theta(NS) = \delta v_\kappa^\infty(0)$. As the one-shot deviation principle (12) specifies, the peer has no incentive to deliberately refuse to upload if $V_\theta(S) \geq V_\theta(NS)$, which can be transformed into the following inequality by taking $M = \lambda b - 1$ [9]

$$\delta(1-\alpha)\left[v_\kappa^\infty(\min\{L, \theta+1\}) - v_\kappa^\infty(0)\right] \geq \lambda bc. \quad (13)$$

---

[5] Since the instant downloading benefit of a peer is not affected by the choice of its strategy. It can be subtracted from the formulation of social norm equilibrium without affecting the analysis.

[6] It should be noted that starting from a particular transaction, the probability that a peer's reputation is increased by complying with the social norm is always larger than $p_\kappa(\min\{L, \theta+1\} \mid \theta)$. Hence, the peer will still have sufficient incentive to follow the social norm if (11) holds.

In the second case when $\tilde{\theta} < h_o$, $\theta$-peers should refuse to upload by complying with the social norm. Thus, the social norm should provide $\theta$-peers incentives to choose $a = NS$ over $a = S$. Similar to (13), the resulting inequality for the peer to have no incentive to upload deliberately is

$$\delta(1-\alpha)\left[v_\kappa^\infty(\min\{L, \theta+1\}) - v_\kappa^\infty(0)\right] \geq -c. \quad (14)$$

*B. Design problem of optimal sustainable social norms*

Based on the above discussion, a protocol can be designed by selecting three parameters: the punishment length $L$, the service threshold $h_o$, and the maximal number of concurrent connections $b$. We assume that the protocol designer aims to choose a protocol that maximizes the social utility (i.e. sharing level among peers) among the candidate protocols that can be sustained as social norm equilibria, then the problem of designing the optimal protocol in this paper can be formalized as follows (we call this problem "optimal social norm equilibrium - OSNE")

---

[7] If an active peer deviates, it will be punished with probability 1. Hence, it has no incentive to comply with the protocol in the subsequent transactions within this period, since its reputation cannot be increased.

[8] It does not affect the analysis when the reputation does not fall to 0 as in a general reputation mechanism, though peers will have a different form of incentive.

[9] If the following inequality holds, $V_\theta(S) \geq V_\theta(NS)$ also holds for a general $M \leq \lambda b - 1$.

$$\begin{aligned}
\underset{(L,h_o,b)}{\text{maximize}} \quad & U_\kappa = \lambda b \mu_\kappa \left[(1-\varepsilon)r - c\right] \\
\text{subject to} \quad & \delta(1-\alpha)\left[v_\kappa^\infty\left(\min\{\theta+1, L\}\right) - v_\kappa^\infty(0)\right] \geq \lambda b c, \ \forall \theta \geq h_o, \quad \text{(OSNE)} \\
& \delta(1-\alpha)\left[v_\kappa^\infty\left(\min\{\theta+1, L\}\right) - v_\kappa^\infty(0)\right] \geq -c, \ \forall \theta < h_o.
\end{aligned}$$

We have proved in [22] that the optimal social utility that can be achieved for a social norm equilibrium always monotonically increases with $L$. Therefore, in the following discussion, we consider the design problem of $(h_o, b)$ given a value of $L$ which can be selected based on the desired complexity of the protocol.

*C. Designing and characterizing the optimal social norm equilibrium*

In this section, we explicitly analyze how the design parameters $(h_o, b)$ will impact the social utility as well as the peers' incentives to comply with the prescribed protocol. This analysis enables us to characterize the optimal design, denoted as $\left(h_o^*, b^*\right)$, which maximizes the social utility while providing peers sufficient incentive to comply with the protocol.

First, we analyze the relationship between the social utility $U_\kappa$ and $(h_o, b)$. We can verify from Problem (OSNE) that $U_\kappa$ monotonically increases with $b$. On the other hand, since $\mu_\kappa = \dfrac{1}{1+\alpha h_o}$ monotonically decreases with $h_o$, we can also conclude that $U_\kappa$ monotonically decreases with $h_o$. Therefore, when we design protocols in a network where peers comply with the protocol, it is always optimal to select $h_o = 1$ and the largest $b$ that is allowed by the system constraints (e.g. the device constraints of peers). Hence, the design problem now becomes selecting the smallest $h_o$ and the largest $b$ for which the incentive constraints in Problem (OSNE) are satisfied.

We then discuss the influence of $(h_o, b)$ on peers' incentives. In particular, we provide the following proposition to establish what conditions should $(h_o, b)$ fulfil (i.e. how should these parameters be selected by the protocol designer) in order to sustain the resulting protocol as a social norm equilibrium.

**Proposition 1.** A protocol $\kappa = (\sigma_o, \tau)$ can be sustained as a social norm equilibrium if and only if

(1) its service threshold $h_o$ is larger than a constant $H_o$ that is defined as

$$h_o \geq H_o \triangleq \ln\left[1 - \frac{(1-\delta)c}{(1-\alpha)[(1-\varepsilon)r-c] - \delta\alpha c}\right] / \ln \delta; \quad (16)$$

(2) the maximum number of concurrent connections $b$ is smaller than a constant $B$, which is the solution of the following equation set

$$\delta(1-\alpha)\left[1-\delta^{h_o}\right]\frac{[(1-\varepsilon)r-c]}{1-\delta(1-\alpha)-\alpha\delta^{h_o+1}}=c. \quad (17)$$

$$\alpha = 1-(1-\varepsilon)^{\lambda B}$$

*Proof*: See Appendix A. ■

Proposition 1 provides a guideline for selecting the parameters $(h_o, b)$ of a P2P reciprocation protocol which can be sustained as a social norm equilibrium. As the proof shows, increasing the service threshold $h_o$ enlarges the gap between the overall utility that can be received by active and inactive peers. Hence, larger values of $h_o$ provide a larger threat of future punishment and thus, provide peers increased incentives to comply with the prescribed protocol. On the other hand, increasing the value of $b$ raises the one-period utility of an active peer and hence enlarges $v_\kappa^\infty(\min\{L,\theta+1\}) - v_\kappa^\infty(0)$. However, larger $b$ also leads to the increase of $\alpha$, since more transactions in one period raises the chance for a peer to make a mistake and thus being punished. The computation in Appendix A shows that $\alpha$ has a more significant impact on peers' incentives than $v_\kappa^\infty(\min\{L,\theta+1\}) - v_\kappa^\infty(0)$ does, and hence, an increasing $b$ reduces the peers' incentives to comply in general. In Section V, we show the trade-off between an increased social efficiency and a decreased incentive to comply with the prescribed protocol by adjusting $h_o$ and $b$.

As mentioned above, there are always practical constraints that need to be taken into consideration when we design the protocol. Here, we consider two specific constraints as $1 \leq h_o \leq L$ and $b > 0$. Failure to find a $(h_o, b)$ within this required region means that there exists no protocol which peers will follow out of their self-interests (i.e. only non-cooperative behaviors can be sustained as a social norm-equilibrium) and thus, the network will collapse in such network settings. In the following proposition, we establish sufficient and necessary conditions for the existence of social norm equilibrium.

**Proposition 2.** (1) There exist protocols that can be sustained as social norm equilibria if and only if

$$\frac{c}{r} \leq T_c \triangleq \frac{\delta(1-\varepsilon)^2\left[1-\delta^L\right]}{1-\delta+\delta\left[1-\delta^L\right]}; \quad (18)$$

(2) There exist protocols that can be sustained as social norm equilibria if and only if a peer's discount factor $\delta$ is larger than or equal to a certain threshold $T_\delta$, which can be determined as

$$T_\delta(1-\varepsilon)\left[1-T_\delta^L\right]\frac{[(1-\varepsilon)r-c]}{1-T_\delta(1-\varepsilon)-\varepsilon T_\delta^{L+1}}=c. \quad (19)$$

*Proof*: See Appendix A. ■

Proposition 2 shows that a non-trivial (cooperative) social norm equilibrium exists if and only if the service cost to benefit ratio is sufficiently small or peers are sufficiently patient. In both cases, peers will put sufficient weight on the future reward which they will obtain from downloads rather than on saving the instant upload cost. Therefore, Proposition 2 can be used as a guideline for designing the optimal social norm by the protocol designer based on its evaluation of the network conditions and peers' characteristics. If $c/r > T_c$ or $\delta < T_\delta$, the network will collapse (i.e. no social norm equilibrium can be sustained in this network). Hence, the P2P system should be redesigned to either decrease the service cost per transaction or increase the peers' patience (the applications need to adopt a larger discount factor).

Based on Proposition 1 and 2, we are able to design the algorithm to determine the optimal values of $(h_o, b)$, denoted as $(h_o^*, b^*)$. The procedure for determining the optimal protocol design algorithm is summarized in Table 3.

*D. Reputation schemes with less harsh punishments*

So far we have focused on reputation schemes under which any deviation from the prescribed protocol is punished with the harshest punishment, i.e. their reputation is reduced to the minimum. Although this class of reputation schemes is simple because the protocol designer needs only to consider one parameter, the number of reputations $L$, such protocols may not yield the highest social utility among all possible reputation schemes when there are service errors. In this section, we discuss less harsh punishments by assuming that, upon deviation, each peer's reputation falls to 0 with a probability less than 1. Particularly, we introduce a probability $\beta \leq 1$. When a $\theta$-peer deviates from the protocol in one period, its reputation falls to 0 with a probability $1 - \beta^{L-\theta+1}$, and remains unchanged with probability $\beta^{L-\theta+1}$. Therefore, the higher reputation a peer has, the larger will its probability to be forgiven upon deviation be. This class of reputation schemes can be identified by two design parameters $(L, \beta)$. (The reputation scheme discussed in Section III can be considered as a special case where $\beta = 0$.) $\beta$ affects the evolution of the reputation distribution, and the stationary distribution of reputations with the reputation scheme $(L, \beta)$ satisfies the following set of equations:

$$\begin{aligned}
\eta_{L,\beta}(L) &= (1 - \alpha + \alpha\beta)\eta_{L,\beta}(L) + (1-\alpha)\eta_{L,\beta}(L-1) \\
\eta_{L,\beta}(\theta) &= \alpha\beta^{L-\theta+1}\eta_{L,\beta}(\theta) + (1-\alpha)\eta_{L,\beta}(\theta-1),\ h_o + 1 \leq \theta \leq L-1 \\
\eta_{L,\beta}(h_o) &= \alpha\beta^{L-h_o+1}\eta_{L,\beta}(h_o) + \eta_{L,\beta}(h_o - 1),\ 1 \leq \theta \leq h_o \\
\eta_{L,\beta}(\theta) &= \eta_{L,\beta}(\theta - 1),\ 1 \leq \theta \leq h_o \\
\eta_{L,\beta}(0) &= \sum_{\theta=h_o}^{L} \alpha\left(1 - \beta^{L-\theta+1}\right)\eta_{L,\beta}(\theta) \\
\mu_{L,\beta} &= \sum_{\theta=h_o}^{L} \eta_{L,\beta}(\theta)
\end{aligned} \quad (20)$$

Let the social utility now denoted as $U_{L,\beta}$, which is still as defined in (10). The protocol designer's problem becomes the following (we call this problem "optimal social norm equilibrium with variable punishment – OSNE/VP")

$$\begin{aligned}\underset{(L,h_o,b,\beta)}{\text{maximize}} \quad & U_{L,\beta} = \lambda b \mu_{L,\beta}\left[(1-\varepsilon)r - c\right] \\ \text{subject to} \quad & \delta(1-\alpha)\left[v_{L,\beta}^{\infty}\left(\min\{\theta+1,L\}\right) - v_{L,\beta}^{\infty}(0)\right] \geq \lambda bc, \ \forall \theta \geq h_o, \quad \text{(OSNE/VP)}\\ & \delta(1-\alpha)\left[v_{L,\beta}^{\infty}\left(\min\{\theta+1,L\}\right) - v_{L,\beta}^{\infty}(0)\right] \geq -c, \ \forall \theta < h_o.\end{aligned}$$

Besides the design of parameters $(L,h_o,b)$, the protocol designer also has to consider the selection of parameter $\beta$ in order to maximize the social utility without violating the incentive constraints. Next, we investigate how $\beta$ impacts the social utility and peers' incentives. First, it can be determined from (10) that $U_{L,\beta}$ monotonically increases with $\mu_{L,\beta}$. That is, the social utility increases as the fraction of peers who can receive upload services increases according to the protocol. As $\mu_{L,\beta}$ monotonically increases with $\beta$, we have the following conclusion.

**Proposition 3.** The social utility $U_{L,\beta}$ monotonically increases with $\beta$.

*Proof*: See Appendix B.

According to Proposition 3, the protocol designer prefers to select $\beta$ as large as possible in order to maximize the efficiency of the network. However, similarly to Problem (OSNE), such a selection is restricted by the incentive constraints for peers to comply with the resulting protocol.

Next, we study $\beta$'s impact on peers' incentives. Similar to the proof of Proposition 1 and 2, we show that peers' incentive to follow a particular social strategy monotonically decreases with $\beta$.

**Proposition 4.** There is a threshold $\beta_\sigma$ for each social strategy $\sigma$ such that $\sigma$ can be sustained in a social norm equilibrium if and only if $\beta \leq \beta_\sigma$

*Proof*: See Appendix B. ∎

We can thus conclude from Proposition 4 that a larger probability $\beta$ to remain at the current reputation for a peer reduces the threat of future punishment, which in turn decreases a peer's incentive to comply with the protocol.

Based on Proposition 3 and 4, it can be concluded that there exist a trade-off between an increased social efficiency and an decreased incentive to comply with the prescribed protocol by adjusting $\beta$. The optimal $\beta$ can be determined using similar algorithms as in Table 3.

*E. Social strategy with various service thresholds*

The current social strategy adopted has a uniform service threshold $h_o$ for all active peers. In this section, we discuss how the performance of a protocol changes when active peers of different reputations have different service thresholds. Particularly, we consider strategies $\sigma \in \Gamma_1$ that can be characterized by a set of service thresholds $h(\sigma) \in \{1, \cdots, L\}$ and $m_\sigma = \{m_\sigma(\theta)\}_{\theta=h(\sigma)}^{L}$ [10], which can be specified as follows

$$\sigma(\theta, \tilde{\theta}) = \begin{cases} S & \text{if } \theta \geq h(\sigma) \text{ and } \tilde{\theta} \geq m_\sigma(\theta) \\ NS & \text{otherwise} \end{cases}, \quad (22)$$

and

$$m_\sigma(h(\sigma)) \leq m_\sigma(h(\sigma)+1) \leq \cdots \leq m_\sigma(L) \text{ for any } \theta \geq h(\sigma). \quad (23)$$

Hence by adopting $\sigma$, an active peer of reputation $\theta \geq h(\sigma)$ will help peers of reputation $\tilde{\theta} \geq m_\sigma(\theta)$. Similar to Section II, we denote the prescribed service thresholds of the social strategy are denoted as $h_o$ and $m_o = \{m_o(\theta)\}_{\theta=h_o}^{L}$.

Now each peer does not provide homogeneous services to others within one period. It can be specified from (23) that active peers with lower reputation provide more services than active peers with higher reputation. Hence, the amount of chunks uploaded by an active peer depends on its reputation $\theta$, the set of service thresholds $m_o$, the reputation distribution $\eta$, as well as $\lambda$ and $b$. Correspondingly, the average service cost consumed by an active peer in one period can be denoted as $q_\kappa(\lambda, b, \theta, m_o, \eta)$ and its expected one period utility being denoted as

$$v_\kappa(\theta) = \lambda b(1-\varepsilon)rI(m_o(h) \leq \theta) - q_\kappa(\lambda, b, \theta, m_o, \eta). \quad (24)$$

The indication function $I(m_o(h) \leq \theta)$ takes the value of 1 if $m_o(h) \leq \theta$, i.e. there exist peers in the network who would help $\theta$-peers according to the prescribed protocol; and it takes the value of 0 if $m_o(h) > \theta$, i.e. there exists no peer who would provide services to $\theta$-peers. The social utility is still defined as the average utility of all peers when the reputation distribution is stationary as follows

$$U_\kappa = \sum_\theta \eta_\kappa(\theta) v_\kappa(\theta) = \sum_{\theta \geq h_o} \eta_\kappa(\theta) \left[ \lambda b(1-\varepsilon)rI(m_o(h) \leq \theta) - q_\kappa(\lambda, b, \theta, m_o, \eta) \right]. \quad (25)$$

Hence, the optimal design problem (OSNE/VP) can be rewritten as follows (we call this problem "optimal social norm equilibrium with variable punishment and service threshold – OSNE/VPS")

---

[10] Here the strategies with the service threshold being $0$ and $L+1$ are not considered.

$$\begin{aligned}
&\underset{(L,h_o,m_o,\beta,b)}{\text{maximize}} && U_\kappa = \sum_{\theta \geq h_o} \eta_\kappa(\theta)\big[\lambda b(1-\varepsilon)rI(m_o(h) \leq \theta) - q_\kappa(\lambda,b,\theta,m_o,\eta)\big] \\
&\text{subject to} && \delta(1-\alpha)\big[v_\kappa^\infty(\min\{\theta+1,L\}) - v_\kappa^\infty(0)\big] \geq \lambda bc, \ \forall \theta \geq h_o, \quad \text{(OSNE/VPS)} \\
& && \delta(1-\alpha)\big[v_\kappa^\infty(\min\{\theta+1,L\}) - v_\kappa^\infty(0)\big] \geq -c, \ \forall \theta < h_o.
\end{aligned}$$

In the following proposition, we characterize how $m_o$ impact the social utility and peers' incentives. In particular, we can prove that (1) the solution for $m_o$ which maximizes the social utility (25) preserves the property as: $h_o = m_o(h_o) \leq m_\sigma(h_o+1) \leq \cdots \leq m_o(L)$ for any $\theta \geq h_o$; (2) the solution for $m_o$ which maximizes peers' incentives to follow the resulting protocol is the one specified as: $m_o(h_o) = h_o$ and $m_o(\theta) = h_o + 1$ for any $\theta \geq h_o + 1$.

**Proposition 5.** (1) The service thresholds $m_o$ which satisfies $m_o(h_o) = h_o$ maximizes the social utility (25) among all possible $m_\sigma$ for $\sigma \in \Gamma_1$.

(2) the set of service thresholds $m_o$ that is defined as follows

$$m_o(\theta) = h_o + 1 \text{ for any } \theta \geq h_o \qquad (27)$$

maximizes peers' incentives. That is, if there exists a set of service thresholds $m_o'$ that satisfies the incentive constraints in Problem (OSNE/VPS), then $m_o$ also satisfies these incentive constraints.

*Proof*: The proof can be conducted from Proposition 1 and 2, which is omitted due to space limitation. ∎

From Proposition 5, we could determine that the optimal solution of Problem (OSNE/VPS) should always set $m_o(\theta) \leq h_o + 1$ for any $\theta \geq h_o + 1$. However, similar to $\beta$, there also exists a trade-off between an increased social efficiency and an decreased incentive to comply with the prescribed protocol by adjusting $m_o(h_o)$ between $h_o$ and $h_o + 1$.

## IV. Protocol Designs for Networks with Altruistic and Free-riding Peers

The analysis in this paper so far assumes that all peers in the network are self-interested and strategic, i.e. they select their actions in order to maximize their long-term utilities. We refer to these peers as "reciprocative" since they will provide services if recipients of the services are likely to return the favour [23]. Nevertheless, in practical P2P networks, there are also other types of peers who may not be strategic and will play the same action constantly. For example in P2P multimedia sharing applications, the protocol designer may usually be able to deploy some peers in the network, e.g. seeds, that have the entire media file, which are called altruistic peers. Altruistic peers provide upload services in response to any request it receives, regardless of the peer's reputation where the request comes from. Meanwhile, they do not send any chunk requests to others. However, due to the bandwidth limitation, the number of upload services that can be provided by an altruistic peer in one period is limited. For the purpose of illustration, we assume that an altruistic peer can provide a maximum number of $\lambda b$ services per period [11]. Hence, an altruistic peer always plays $a = 1$ in any upload transaction in which it is engaged. In this way, the protocol designer tries to increase the sharing efficiency in the network by adjusting the fraction of altruistic peers in the peer population, denoted by $p_C$. Since altruistic peers are deployed by the protocol designer, we assume that they can be identified by the system. That is, an altruistic peer will be assigned a reputation $L$ by the protocol constantly, regardless of whether its upload is successful or not.

On the contrary, there are also malicious peers, whose goal is to cause damages to other peers and attack the system. The most common attacks include incomplete chunk attack and pollution attack [24]. In an incomplete chunk attack, a malicious peer agrees to send the entire requested chunk to its client, but sends only portions of it or no data at all. In a pollution attack, a malicious peer corrupts the media chunks, renders the content unreadable, and then makes this polluted content available for sharing with other peers. In both cases, the client of a malicious peer wastes its download connection and has to request the same chunk again in a separate transaction. Meanwhile, a malicious peer is regarded to be playing $a = 0$ in any upload transaction it is engaged. The fraction of malicious peers in the population is denoted as $p_D$. Here we assume that malicious peers are treated by the protocol as regular reciprocative peers, whose reputation will increase after complying with the protocol and decrease after deviating from the protocol.

In this section, we analyze the impact of such altruistic and free-riding population on the design of P2P protocols. The fraction of reciprocative peers is correspondingly denoted as $p_R = 1 - p_C - p_D$. First, we analyze the impact of malicious peers by assuming $p_C = 0$, i.e. $p_R = 1 - p_D$ [12]. To simplify the analysis, we consider the reputation scheme with $\beta = 0$ and the social strategy with $m_o(\theta) = h_o$ for all $\theta \geq h_o$, but a similar analysis can be performed for other schemes and social stategies. Similar to the previous sections, we are also interested in the long run stationary distribution of reputation, denoted as $\{\eta_D(\theta)\}_{\theta=0}^{L}$, with $\mu_D$ denoting the fraction of peers that can receive services according to the protocol. Since a malicious peer does not provide any complete or uncorrupted chunks to other peers, its reputation will follow the iterative process of being increased by 1 per period first until reaching $h_o$, and then falling to $0$. Let $\{\omega_D(\theta)\}_{\theta=0}^{L}$ denote the stationary distribution of the malicious peer population, which can be expressed as follows:

---

[11] The results in this section do not change when the maximum number of services is not $\lambda b$.

[12] The impacts of malicious peer and altruistic peer are additive. Hence, the results in this section can be applied to networks where malicious peers and altruistic peers co-exist.

$$\begin{aligned}
&\omega_D(\theta) = 0, \ h_o+1 \leq \theta \leq L \\
&\omega_D(\theta) = \omega_D(\theta-1), \ 1 \leq \theta \leq h_o . \\
&\omega_D(0) = \omega_D(h_o)
\end{aligned} \quad (28)$$

Meanwhile, let $\{\omega_R(\theta)\}_{\theta=0}^{L}$ denote the stationary distribution of reciprocative peers, and it can be computed using (8). Summing up, the stationary reputation distribution of the population can be solved using the following set of equations:

$$\begin{aligned}
&\eta_D(\theta) = (1-p_D)\omega_R(\theta) + p_D\omega_D(\theta), \ 0 \leq \theta \leq L \\
&\mu_D = \sum_{\theta=h_o}^{L} \eta_D(\theta)
\end{aligned} \quad (29)$$

Since malicious peers consume a reciprocative peer's download rate, the reciprocative peer tries to download $\lambda b \dfrac{p_D}{\mu_D + \dfrac{h}{h+1}p_D}$ chunks per period from malicious peers while downloading the other

$\lambda b \dfrac{\mu_D - \dfrac{1}{h+1}p_D}{\mu_D + \dfrac{h}{h+1}p_D}$ chunks from other reciprocative peers. Therefore, an active reciprocative peer's

expected one-period utility now can be formalized as

$$v_\kappa(\theta) = \lambda b \dfrac{\mu_D - \dfrac{1}{h+1}p_D}{\mu_D + \dfrac{h}{h+1}p_D}[(1-\varepsilon)r - c], \text{ for } \theta \geq h_o, \quad (30)$$

which monotonically decreases with $p_D$ [13].

From (30), we can thus determine that the average number of chunks reciprocative peers can receive monotonically decreases with $p_D$ and so is their average one-period utility. The stationary distribution (29) and the expected one-period utility (30) also enable us to analyze the reciprocative peers' incentives to comply with the protocol using one-shot deviation principle. The result is formalized in the following proposition.

**Proposition 6.** Given a protocol $\kappa = (\sigma_o, \tau)$ and the fraction of malicious peers $p_D$, peers' incentive to comply with $\kappa$ monotonically decrease with $p_D$.

*Proof*: See Appendix C. ∎

---

[13] Here we assume that a reciprocative peer knows the value of $p_D$ by measuring how many incomplete or polluted chunks it receives in the past.

Proposition 6 indicates that the presence of malicious peers does not only decrease the social welfare, but also their incentives to comply with the protocol.

Next, we investigate how altruistic peers impact the reciprocative peers' utilities and incentives with $p_D = 0$ and $p_R = 1 - p_C$. Similarly, we let $\{\eta_C(\theta)\}_{\theta=0}^{L}$ denote the corresponding stationary distribution and $\mu_C$ denote the fraction of peers that can receive services according to the protocol.

Since an altruistic peer is assigned a constant reputation of $L$ by the system, $\{\eta_C(\theta)\}_{\theta=0}^{L}$ can be computed as

$$\eta_C(\theta) = (1 - p_C)\omega_R(L) + p_C$$
$$\eta_C(\theta) = (1 - p_C)\omega_R(\theta),\ 0 \leq \theta \leq L. \tag{31}$$
$$\mu_C = \sum_{\theta=h_o}^{L} \eta_C(\theta)$$

Here, as altruistic peers do not download chunks from other peers, the average upload cost of an active reciprocative peer can be calculated as $\dfrac{\lambda b(\mu_C - p_C)}{\mu_C} c$ as some of its upload tasks are shared by altruistic peers. Hence, an active reciprocative peer's expected one period utility now becomes

$$v_\kappa(\theta) = \lambda b(1 - \varepsilon)r - \frac{\lambda b(\mu_C - p_C)}{\mu_C} c,\ \text{for } \theta \geq h_o. \tag{32}$$

An inactive reciprocative peer can also receive services from an altruistic peer. Hence, its expected one-period utility now becomes

$$v_\kappa(\theta) = \begin{cases} \lambda b(1 - \varepsilon)\dfrac{p_C}{1 - p_C}r,\ \text{if } \theta < h_o\ \text{and}\ p_C \leq 0.5 \\ \lambda b(1 - \varepsilon)r,\ \text{if } \theta < h_o\ \text{and}\ p_C > 0.5 \end{cases}. \tag{33}$$

It is easy to specify from (32) and (33) that altruistic peers reduce the upload cost of active peers while raising the download benefit of inactive peers. Hence, the average utility of reciprocative peers monotonically increases with $p_C$. Also, by incorporating (31) - (33) into the one-shot deviation principle, we show that $p_C$ cannot be too large in order to sustain a reciprocative peer's incentive to comply with the protocol.

**Proposition 7.** Given a protocol $\kappa = (\sigma_o, \tau)$ and the fraction of altruistic peers $p_C$, $\kappa$ can be sustained as a social norm equilibrium if and only if $p_C$ is below certain threshold $\overline{p}_C \leq 0.5$.

*Proof*: See Appendix C. ∎

Proposition 7 provides a counter-intuitive result as it is not always good to increase $p_C$ in the network. Although having more altruistic peers allows more upload services in the network, they in turn harm a peer's incentive to comply with the protocol, which reduces the cooperative sharing behavior among reciprocative peers.

Figure 2 plots the average utility of reciprocative peers in the network against $p_D$ and $p_C$. It shows that the utility monotonically decreases with $p_D$. When $p_D$ reaches certain points, peers lose their incentive to follow the protocol and the network collapses with the average utility falling to 0. Meanwhile, the utility does not monotonically increases with $p_C$, since reciprocative peers lose the incentive to comply with the protocol at certain point of $p_C$. However, as $p_C$ approaches 1, the average utility finally reaches the optimal value $\lambda b (1-\varepsilon) r$ since all peers' download requests can be fully served by altruistic peers.

Although the average utility of reciprocative peers reaches the optimum when $p_C$ approaches 1, the protocol designer cannot choose a $p_C$ that is arbitrarily large due to the fact that the sharing behavior of altruistic peers also incurs upload costs. As an altruistic peer has all the chunks, it cannot gain any benefit by receiving chunks from other peers, we thus regard an altruistic peer's utility in one transaction as its upload cost. Figure 3 plots the social utility of all peers in the network, including altruistic peers and reciprocative peers, against $p_C$. This figure shows that the social utility does not increase monotonically with $p_C$. Therefore, the protocol designer should select the optimal value of $p_C$ as the point that maximizes the social utility of all peers – an optimization which is formalized as follows (we call this problem "optimal social norm equilibrium with altruistic peers – OSNE/AH"):

$$\begin{aligned} \underset{(L, h_o, b, p_C)}{\text{maximize}} \quad & U_\kappa \\ \text{subject to} \quad & \delta(1-\alpha)\big[v_\kappa^\infty\left(\min\{\theta+1, L\}\right) - v_\kappa^\infty(0)\big] \geq \lambda b c, \ \forall \theta \geq h_o, \quad \text{(OSNE/AH)} \\ & \delta(1-\alpha)\big[v_\kappa^\infty\left(\min\{\theta+1, L\}\right) - v_\kappa^\infty(0)\big] \geq -c, \ \forall \theta < h_o \end{aligned}$$

where

$$U_\kappa = \begin{cases} \lambda b (1-p_C)[(1-\varepsilon)r - c], & \text{if } p_C > 0.5 \\ \lambda b (1-\varepsilon)\left[\dfrac{p_C}{1-p_C}(1-\mu_C) + (\mu_C - p_C)\right]r - \lambda b \left[\dfrac{(\mu_C - p_C)^2}{\mu_C} - p_C\right]c, & \text{if } p_C \leq 0.5 \end{cases} \quad (35)$$

The problem (OSNE/AH) can be solved using a similar iterative algorithm as the one in Table 3. Figure 4 plots the optimal social utility and the corresponding optimal fraction of altruistic peers as $p_C^*$ by solving (OSNE/AH).

# V. ILLUSTRATIVE EXAMPLES

*A. Simulation settings*

In this section, we illustrate the impact of the proposed social norm based protocols on P2P multimedia sharing networks using the simulator built up by Matlab. A number of 200 peers are deployed in the network. All peers have the same download rate of 1Mbps. In each experiment, peers exchange a single video file of approximate size 100 Mbits, at CIF (352×528) resolution and 30 frames per second. The video is encoded using H.264/AVC codec and divided into chunks of 0.1s. All peers join the network at the same time. In the experiments, we deploy a reputation set $\Theta = \{0,1,2,3\}$, i.e. $L = 3$, $\beta = 0$ and $m_o(\theta) = h_o$ for all $\theta \geq h_o$. We keep $\Theta$ and $L$ fixed during the experiments.

*B. Performance evaluation*

*(1) The impact of network conditions*

The performance of a protocol is not only affected by the design parameters $(L, h_o, b)$, but also by the parameters $(r, c, \varepsilon)$, which are intrinsic to the network conditions, as well as $(\lambda, \delta)$, which are selected by peers. In this and the following sections, we discuss how the performances of social norm based protocols are influenced by these intrinsic parameters. We first assess the impact of the network conditions.

From the analysis in Section III, it can be observed that the impact of $r$ and $c$ on peers' incentives is reflected through the service cost to benefit ratio $c/r$. Figure 5 plots the social utility of protocols with $h_o = 1, 2, 3$, respectively, by varying $c/r$ and $\varepsilon$ from 0 to 1. $b$ is optimized based on Problem (OSNE). This figure highlights that the social utility decreases monotonically as the network conditions become worse, since larger $c/r$ and $\varepsilon$ reduce the threat of punishment provided by the protocol. This figure also illustrates the trade-off between social utility and incentives for peers to comply with the protocol. Taking the maximum of the social utilities corresponding to $h_o = 1, 2, 3$ at each point of $c/r$ and $\varepsilon$, the optimal social utility is also plotted in the figures as the green curves, with the corresponding $h_o$ at each point being the optimal service threshold $h_o^*$ that solves Problem (OSNE). Although protocols with low $h_o$ provide higher social utility when the network conditions are good with small $c/r$ and $\varepsilon$, protocols with high $h_o$ provides larger incentives for peers and thus have better performance when the network conditions worse off. Therefore, $h_o^*$ monotonically increases with both $c/r$ and $\varepsilon$. When $c/r > 0.9$ or $\varepsilon > 0.95$, there exists no protocol that can be sustained as social norm equilibrium and the social utility falls to 0.

*(2) The impact of peer characteristics*

We then discuss how the peer characteristics impact the protocols' performance. The results are shown in Figure 6.

With $\delta$ increasing, a peer gives higher weight on its overall utility instead of its instant utility. As Proposition 2 shows, the social norm will be more effective in incentivizing peers to cooperate in such cases. It should also be noted that the protocol with high $h_o$ has a better performance when $\delta$ is small, even though the punishment prohibits the sharing activity between peers. When $\delta$ is between $0.2$ and $0.3$, $L$-peers with $h_o = 2$ and $3$ already have incentives to comply with the protocol, and will mutually cooperate with each other; while $L$-peers with $h_o = 1$ are still disincentivized and will not cooperate. Therefore, $h_o^*$ monotonically decreases with $\delta$.

As $\lambda$ represents the rate that each connection is utilized per period, the social utility in general increases with $\lambda$, since higher frequency of chunk exchange between peers makes the network more socially valuable. However, as a peer engages in more transactions per period, the probability $\alpha$ that it will be falsely punished by the protocol also increases. Hence, a larger $\lambda$ reduces peers' incentives, which are reflected at points when the social utility falls instantly. Therefore, when $\lambda$ is small, it is beneficial to select a large $b$ to increase the chunk exchange frequency. On the other hand, when $\lambda$ is large and the network is congested, selecting a small $b$ is better so as to reduce the error probability. Determined from the optimal social utility plotted as the green curve, the optimal $b^*$ decreases with $\lambda$.

*(3) Video Quality Comparison*

In this section, we explicitly compare the average PSNR of the decoded video among all peers using different protocols. The exchanged video content is the well-known "Foreman" sequence repeated multiple times to create a long sequence. Besides the protocols studied in this paper, the performance of the Tit-for-Tat (TFT) protocol is also analyzed. To make the TFT applicable to networks with random matching features, we slight change the protocol, which is defined as follows.

- The reputation set is binary as $\Theta = \{0,1\}$.

- The social strategy $\sigma_{TFT}$ is defined as: $\sigma_{TFT}(\theta, \tilde{\theta}) = S$ if $\tilde{\theta} = 1$; $\sigma_{TFT}(\theta, \tilde{\theta}) = NS$ if $\tilde{\theta} = 0$.

- Using the same rule to calculate the statistic $x$ as in Section II, $\tau(\theta, x) = 0$, $\forall \theta \in \Theta$. The reputation scheme is defined as: $\tau(\theta, x = 0) = 1$ and $\tau(\theta, x = 1) = 0$.

Table 4 presents the results given the parameters $b = 5$, $\lambda = 1$, and $\delta = 0.8$ for all peers. We also deploy 10% of altruistic peers in the network as seeds, i.e. $p_C = 0.1$. Meanwhile, we also consider variable combinations of $(c/r, \varepsilon)$. There are four strategies being considered: the threshold-based strategy with $h_o = 1, 2, 3$ respectively, and TFT. For the threshold-based strategy, we choose the optimal values of $\beta$ and $m_o$ to maximize Problem (OSNE/VPS). When the service cost to benefit ratio and the service error are low, TFT delivers a PSNR which is comparable to those of threshold-based strategies. Meanwhile, as TFT only has two reputation levels and less peers being falsely punished, its PSNR is higher than those of threshold-based strategies with $h_o = 2, 3$. On the other hand, our social norm based protocols are more robust by using threshold-based strategies, which deliver performances that are more insensitive to the variation on network conditions.

Figure 7 illustrates the advantage of optimal social norm equilibrium over fixed protocols which are selected in ad-hoc by explicitly comparing the performances of the following protocols

- Protocol 1: all peers cooperate unconditionally without considering the incentive constraints.
- Protocol 2: optimal social norm equilibrium with $h_o$, $b$, $\beta$, and $p_C$ being optimized.
- Protocol 3: a fixed social norm with $h_o = 3$, $b = 5$, $\beta = 0$, and $p_C = 0.3$.
- Protocol 4: TFT with $b = 5$, $\beta = 0$ and $p_C = 0.3$.

Since all peers provide full services in Protocol 1, the performance it delivers remains to be constant and serves as the Pareto boundary of the performance that an incentive protocol can possibly achieve. Using this as a benchmark, Figure 7 shows that the optimal social norm equilibrium leads to significant improvements in terms of PSNR over Protocol 3 and 4, both of which adopt fixed strategies. As the PSNR delivered by the optimal social norm equilibrium remains roughly constant against the variation of $c/r$, the PSNR delivered by Protocol 3 and 4 drastically decrease with $c/r$. When $c/r$ exceeds 0.25, the network adopting TFT collapses; while such collapse also happens in the network adopting Protocol 3 when $c/r$ exceeds 0.45. In both cases, the reciprocative peers lose their incentive to follow the protocols and do not mutually provide upload services at all. Hence, there only exist minimum upload services in the network which are provided by the altruistic peers.

## VI. POSSIBLE EXTENSIONS AND FUTURE RESEARCH

In this section we discuss possible extensions to our proposed framework to accommodate some other unique features of P2P multimedia sharing applications.

### A. Scalable video delivery

Our proposed framework in Section II assumes that all the chunks are of equal size and they have the same benefit (value) in terms of the multimedia distortion reduction. This framework can be improved by explicitly considering the priorities of the various chunks. The set of priority classes may depend on the specific video encoder used by the content creator. Using standard-based video codecs (e.g. H.264/AVC or MPEG-2) as the example, video files are typically compressed into three classes of frames (Intra (I), Predictive (P), and Bi-directionally predictive (B)). In addition, each frame can be classified by its activity level taking values from the set {*High, Medium, Low*} in order to capture the variations in activity level (e.g. motion) between scenes [19]. Assuming that chunks are partitioned into $J$ classes under this priority classification model, each chunk from class $j \in \{1, 2, \cdots, J\}$ has a class-dependent value $r_j$ and a size $d_j$.

For example, a chunk containing high-motion content and/or I-frames should have a higher value and a larger size than a chunk containing low-motion content and/or B-frames. Moreover, the upload cost and the service error per transaction also vary according to the size of the chunk, denoted as $c(d_j)$ and $\varepsilon(d_j)$, respectively. In general, we should have that $c(d_j) > c(d_j')$ and $\varepsilon(d_j) > \varepsilon(d_j')$ for any $d_j > d_j'$, implying that a chunk with larger size should incur a higher upload cost to the server and a higher probability to be lost during transmission.

As the chunks that a peer uploads in different transactions are no longer homogeneous, the calculation of a peer's contribution in one period should put different weights on transactions depending on the properties of chunks uploaded in order to differentiate among the priority classes. In particular, the contribution, denoted as $y$, is measured as the total value of chunks that a peer uploads in one period, that is, $y = \sum_{i=1}^{\lambda b} r^i x^i$, where $r^i$ denotes the chunk value and $x^i \in \{0,1\}$ denotes the output of each transaction respectively for the peer's $i$-th upload transaction in one period. Due to the fact that $r^i$ takes a finite number of values, $y$ also takes values from a finite set, denoted as $Y$. Hence, $\tau$ determines the peer's new reputation as $\tau : \Theta \times Y \to \Theta$. In this way, the reputation scheme updates the reputation depending not only on whether a peer contributes or not, but also on how much a peer contributes. Therefore, the incentives for peers to cooperate and contribute can be further stimulated, which can significantly increase the efficiency of the protocol. Since both $\Theta$ and $Y$ are finite, the reputation scheme can be represented as a lookup table in practical implementations. An illustrative example of such lookup table is presented in Table 5 by setting $L = 5$, $\lambda = 0.5$, $b = 2$, $J = 3$, and $r_j \in \{5, 10, 20\}$, where each entry represents the new reputation after the update.

Moreover, the priority classification of media chunks also affects a peer's incentive to comply with the protocol, as different $c(d_j)$ and $\varepsilon(d_j)$ according to different chunk sizes impact a peer's evaluation on its instant utility and thus influence the strength of threat from future punishments that are imposed by the protocol. Specifically, as a larger chunk incurs higher upload cost in the current transaction as well as a higher probability of service error which in turn reduces its expected future utility, the protocol might be unable to provide sufficient incentives for a peer who is willing to upload a short chunk to comply with. In order to design a robust protocol in this case, one solution would be to select sufficiently large $h_o$ and sufficiently small $b$ such that the protocol can still provide incentives for peers to comply with for the smallest value in $\{c(d_j)\}_{j=1}^{J}$ and the largest value of $\{\varepsilon(d_j)\}_{j=1}^{J}$. However, such design might not be efficient since large $h_o$ and small $b$ will introduce strong punishment in the network and thus reduce the social utility. Therefore by adjusting $h_o$ and $b$, we could also propose an alternative approach to design a protocol which might not be in equilibrium for all possible chunk types, but only ensures peers' incentives to upload chunks types which are most important to the video quality (e.g. chunks containing I-frames). With a lower level of punishment than in the previous approach, the social utility of the network can be possibly raised.

*B. Whitewashing*

In our proposed framework, we do not consider the population dynamics in P2P networks where peers leave and join the network dynamically. If peers newly joining the network are treated undiscriminatively by assigning them the same reputation, an existing peer in the network with a low reputation might deliberately leave and rejoin the network in order to acquire a new identity and a higher reputation. This is commonly known as the whitewashing effect [14]. To address this problem, a straightforward approach is to assign each peer in the network a fixed identity (e.g. a username) which cannot change over time. However, this incurs a high implementation cost which hinders the scalability of the P2P system. Meanwhile, a peer can still take advantage by registering multiple identities. Under our framework, the whitewashing effect can be effectively mitigated by adjusting the initial reputation that is assigned to newly joined peers. Since a peer's incentive to whitewash depends on its comparison between the instant cost it expenses on leaving and rejoining the network and the increase on future utility it can gain with a better new reputation, we could set the initial reputation to the highest value that is just enough to provide the peer an increase on future utility that is smaller than its whitewashing cost.

*C. False reporting*

We assume clients report truthfully to the tracker about the services received from servers. However, peers will have incentives to lie in their reports if this brings them positive effects on their long-term utilities by either raising their own reputations or reducing the reputations of their opponents. The reason why false reporting can potentially subvert our social norm based protocols is that a peer's reputation puts equal weights on the opinion of everyone in the system, which is called "*objective reputation*". To address this problem, a peer can compute the reputations of its opponents subjectively by incorporating its own opinions based on its own interactions with specific peers, which is called "*subjective reputation*". The peer can then combine the subjective reputation and the objective reputations to evaluate a particular peer. Depending on how much the peer trusts this opponent (e.g. the interaction frequency between them, the topological connections, etc.), the peer can dynamically adjust the weights it puts to the subjective and objective reputations, respectively. Under this approach, the reliability of our reputation scheme increases as peers in general put larger weights on the subjective reputation when they meet an opponent with which they had frequent interactions with and thus they are familiar with, while they put larger weights on the objective reputation when they meet a stranger.

## VII. CONCLUSIONS

In this paper, we build on our theoretical framework in [22] to analyze and design the incentive protocols based on indirect reciprocity for P2P multimedia sharing applications. We designed optimal social norms which are sustainable and thus, under which no peer gains by deviating from the prescribed social strategy and thus have no incentive to deviate deliberately. We investigated the design of optimal incentive protocols in order to maximize the sharing efficiency of the network. We analyzed the structures of optimal incentive protocols, identifying the trade-off between efficiency and incentives, and proposed a simple protocol design algorithm. We also discussed the impact of variable punishment, variable service thresholds, as well as altruistic and malicious populations, on the design and performance of optimal incentive protocols. Our simulation results illustrate the impacts of the network conditions and peer characteristics on the performance of incentive protocols and verify that our social norm based protocol can deliver better performance than traditional incentive protocols. Lastly, as the future research, we discussed possible extensions on our proposed framework which consider other unique features of P2P multimedia sharing applications.

## APPENDIX A

### PROOF OF PROPOSITION 1

From (5) and (6), the expected overall utilities can be represented recursively as follows:

$$v_\kappa^\infty(\theta) = v_\kappa(\theta) + \delta\big[(1-\alpha)v_\kappa^\infty(\min\{L,\theta+1\}) + \alpha v_\kappa^\infty(0)\big], \text{ for } \theta \geq h_o; \tag{37}$$

and

$$v_\kappa^\infty(\theta) = \delta v_\kappa^\infty(\min\{L,\theta+1\}), \text{ for } \theta < h_o. \tag{38}$$

Substituting (3) into (37) and (38), it is easy to specify that $v_\kappa^\infty(\theta)$ is non-decreasing with $\theta$, i.e. $v_\kappa^\infty(\theta_1) \geq v_\kappa^\infty(\theta_2)$ when $\theta_1 > \theta_2$.

From (37), it can be derived that for any $\theta \geq h_o$,

$$v_\kappa^\infty(\theta+1) - v_\kappa^\infty(\theta) = \delta(1-\alpha)\big[v_\kappa^\infty(\min\{L,\theta+2\}) - v_\kappa^\infty(\min\{L,\theta+1\})\big]. \tag{39}$$

Particularly,

$$v_\kappa^\infty(\theta+1) - v_\kappa^\infty(\theta) = \delta(1-\alpha)\big[v_\kappa^\infty(\theta+2) - v_\kappa^\infty(\theta+1)\big], \text{ for } h_s \leq \theta \leq L-1, \tag{40}$$

$$\text{and } v_\kappa^\infty(L) - v_\kappa^\infty(L-1) = \delta(1-\alpha)\big[v_\kappa^\infty(L) - v_\kappa^\infty(L)\big] = 0. \tag{41}$$

Therefore, we have the conclusion that $v_\kappa^\infty(\theta) = v_\kappa^\infty(h_o)$ for all $\theta > h_o$. Substituting this into (37), we have that

$$v_\kappa^\infty(h_o) = \lambda b\left[(1-\varepsilon)r - c\right] + \delta\left[(1-\alpha)v_\kappa^\infty(h_o) + \alpha\delta^{h_o}v_\kappa^\infty(h_o)\right]. \tag{42}$$

Hence, $v_\kappa^\infty(h_o)$ is solved as

$$v_\kappa^\infty(h_o) = \frac{\lambda b\left[(1-\varepsilon)r - c\right]}{1 - \delta(1-\alpha) - \alpha\delta^{h_o+1}}. \tag{43}$$

Due to the monotonicity of $v_\kappa^\infty(\theta)$, only two incentive constraints from Problem (OSNE) need to be checked in order to determine a protocol's equilibrium property. For reputations smaller than $h_o$, it has to be verified that

$$\delta(1-\alpha)\left[v_\kappa^\infty(1) - v_\kappa^\infty(0)\right] = \delta(1-\alpha)\left(\delta^{h_o-1} - \delta^{h_o}\right)v_\kappa^\infty(h_o) \geq -c, \tag{44}$$

and for reputations larger than or equal to $h_o$, it has to be verified that

$$\delta(1-\alpha)\left[v_\kappa^\infty(h_o+1) - v_\kappa^\infty(0)\right] = \delta(1-\alpha)\left(1 - \delta^{h_o}\right)v_\kappa^\infty(h_o) \geq \lambda bc. \tag{45}$$

Once (44) and (45) are satisfied at the same time, we can thus conclude that the protocol $\kappa$ is a social norm equilibrium.

Substituting (43) into (44) and (45), the incentive constraints for a protocol to be sustained as a social norm equilibrium can be finally written as

$$\delta(1-\alpha)\delta^{h_o-1}[1-\delta]\frac{\lambda b\left[(1-\varepsilon)r - c\right]}{1 - \delta(1-\alpha) - \alpha\delta^{h_o+1}} \geq -c, \tag{46}$$

$$\delta(1-\alpha)\left[1 - \delta^{h_o}\right]\frac{\lambda b\left[(1-\varepsilon)r - c\right]}{1 - \delta(1-\alpha) - \alpha\delta^{h_o+1}} \geq \lambda bc. \tag{47}$$

Since $1 - \delta(1-\alpha) - \alpha\delta^{h_o+1} = 1 - \delta + \delta\alpha\left(1 - \delta^{h_o}\right) > 0$, it can be determined that when $\frac{c}{r} \leq 1 - \varepsilon$, (46) is satisfied. As $r$ is usually large compared to $c$ in P2P multimedia services and $\varepsilon$ is small, we assume that $\frac{c}{r} \leq 1 - \varepsilon$ and hence (46) always holds.

By transforming (47), we have that

$$\delta^{h_o} \leq 1 - \frac{(1-\delta)c}{(1-\alpha)\left[(1-\varepsilon)r - c\right] - \delta\alpha c}. \tag{48}$$

Taking logarithm over both sides, we have that

$$h_o \geq \ln\left[1 - \frac{(1-\delta)c}{(1-\alpha)\left[(1-\varepsilon)r - c\right] - \delta\alpha c}\right] / \ln\delta. \tag{49}$$

In terms of $b$, the problem is more complicated since $\alpha = 1 - (1-\varepsilon)^{\lambda b}$ is also a function of $b$. To analyze how (47) changes with $b$, we only have to determine how the following term changes with $b$

$$\frac{(1-\alpha)}{1-\delta(1-\alpha)-\alpha\delta^{h_o+1}} = \frac{(1-\varepsilon)^{\lambda b}}{1-\delta+\delta\left(1-(1-\varepsilon)^{\lambda b}\right)\left(1-\delta^{h_o}\right)}. \tag{50}$$

Since $(1-\varepsilon)^{\lambda b}$ monotonically decreases with $b$, so is (50). Hence, the left-hand side of (47) is a decreasing function of $b$. Therefore, the value of $b$ should be below certain threshold $B$ in order for (47) to be satisfied. ∎

PROOF OF PROPOSITION 2

As we have shown in Proposition 1, peers' incentives monotonically increases with $h_o$ and monotonically decreases with $b$. Hence, the pair of design parameters that can maximize peers' incentives is $(h_o = L, b = 1)$. If the incentive constraint (47) cannot be satisfied in this case, we can then draw the conclusion that there exists no protocol that can be sustained as a social norm equilibrium. Substituting $(h_o = L, b = 1)$ into (47), we have that

$$\delta(1-\varepsilon)\left[1-\delta^L\right]\frac{[(1-\varepsilon)r-c]}{1-\delta(1-\varepsilon)-\varepsilon\delta^{L+1}} \geq c. \tag{51}$$

Reorganizing (51), we have that

$$\frac{c}{r} \leq \frac{\delta(1-\varepsilon)^2\left[1-\delta^L\right]}{1-\delta+\delta\left[1-\delta^L\right]}. \tag{52}$$

So there exists social norm equilibrium if and only if (52) is satisfied.

Similarly, we can analyze the relationship between the incentives and the discount factor $\delta$. We only have to determine the term $\dfrac{\delta\left[1-\delta^{h_o}\right]}{\delta-\alpha\delta\left(1-\delta^{h_o}\right)}$ changes against $\delta$. Taking derivative, we have this term monotonically increases with $\delta$. ∎

APPENDIX B

PROOF OF PROPOSITION 3

From (20), it can be determined that $\eta_{L,\beta}(0)$ and hence $\eta_{L,\beta}(\theta)$ for $\theta < h_o$ monotonically decrease with $\beta$. Therefore, $\mu_{L,\beta}$ and so is the social utility $U_{L,\beta}$ monotonically increase with $\beta$. ∎

PROOF OF PROPOSITION 4

By complying with the protocol, a peer of reputation $\theta \geq h_o$ will become $\theta + 1$ after one period with probability $(1-\alpha)$, $\theta$ with probability $\alpha\beta^{L-\theta+1}$, and $0$ with probability $\alpha\left(1-\beta^{L-\theta+1}\right)$. By deviating from the protocol, the peer with have reputation $\theta$ with probability $\beta^{L-\theta+1}$ and with reputation $0$ with probability $\left(1-\beta^{L-\theta+1}\right)$. Hence, the incentive constraint now becomes

$$\delta(1-\alpha)\left[v_\kappa^\infty\left(\min\{\theta+1, L\}\right) - \beta^{L-\theta+1}v_\kappa^\infty(\theta) - \left(1-\beta^{L-\theta+1}\right)v_\kappa^\infty(0)\right] \geq c, \text{ for } \theta \geq h_o. \quad (53)$$

The left-hand sides of (53) decreases as $\beta$ increases. Therefore, peers' incentive to follow the protocol monotonically decreases with $\beta$. Hence, there is a threshold $\beta_\sigma$ for each social strategy $\sigma$ such that $\sigma$ can be sustained in a social norm equilibrium if and only if $\beta \leq \beta_\sigma$. ∎

APPENDIX C

PROOF OF PROPOSITION 6

Substituting (30) into (37) and (38), the incentive constraints for a protocol to be sustained as a social norm equilibrium can be written as

$$\delta(1-\alpha)\delta^{h_o-1}[1-\delta]\frac{\lambda b \dfrac{\mu_D - \dfrac{1}{h+1}p_D}{\mu_D + \dfrac{h}{h+1}p_D}[(1-\varepsilon)r - c]}{1-\delta(1-\alpha) - \alpha\delta^{h_o+1}} \geq -c, \quad (54)$$

$$\delta(1-\alpha)\left[1-\delta^{h_o}\right]\frac{\lambda b \dfrac{\mu_D - \dfrac{1}{h+1}p_D}{\mu_D + \dfrac{h}{h+1}p_D}[(1-\varepsilon)r - c]}{1-\delta(1-\alpha) - \alpha\delta^{h_o+1}} \geq \lambda bc. \quad (55)$$

As $\mu_D \geq \dfrac{1}{h+1}p_D$ and $(1-\varepsilon)r > c$, (54) always holds. The left-hand side of (55) monotonically decrease with $p_D$. Hence, if a protocol can be sustained as a social norm equilibrium for some $p_D$, it can also be sustained for any $p_D' < p_D$. ∎

PROOF OF PROPOSITION 7

From (37) and (38), it can be determined that $v_\kappa^\infty(\theta) - v_\kappa^\infty(0)$ monotonically increases with $v_\kappa(\theta) - v_\kappa(0)$ for any $\theta$. Hence, to analyze the impact of $p_C$ to peers' incentives, we only have to analyze how $p_C$ influences $v_\kappa(\theta) - v_\kappa(0)$. Since $v_\kappa(\theta) - v_\kappa(0) = 0$ for $\theta < h_o$, we only have to analyze $v_\kappa(\theta) - v_\kappa(0)$ for $\theta \geq h_o$, which can be written as follows

$$v_\kappa(\theta) - v_\kappa(0) = \begin{cases} \lambda b(1-\varepsilon)\dfrac{1-2p_C}{1-p_C}r - \dfrac{\lambda b(\mu_C - p_C)}{p_C}c, & if\ p_C \leq 0.5 \\ -\dfrac{\lambda b(\mu_C - p_C)}{p_C}c, & if\ p_C > 0.5 \end{cases}, \text{ for } \theta \geq h_o. \quad (56)$$

It can be determined from (56) that there exists a $\tilde{p}_C$ such that $v_\kappa(\theta) - v_\kappa(0)$ monotonically decreases with $p_C$ after $p_C > \tilde{p}_C$. Hence, there also exists a $\tilde{p}_C \leq \overline{p}_C$ such that when $p_C > \overline{p}_C$, incentive constraints in (44) and (45) no longer hold. As when $p_C > 0.5$, $v_\kappa(\theta) - v_\kappa(0) < 0$ for $\theta \geq h_o$ and incentive constraint (45) does not hold. We conclude that $\overline{p}_C \leq 0.5$. ∎

## APPENDIX E

## FIGURES

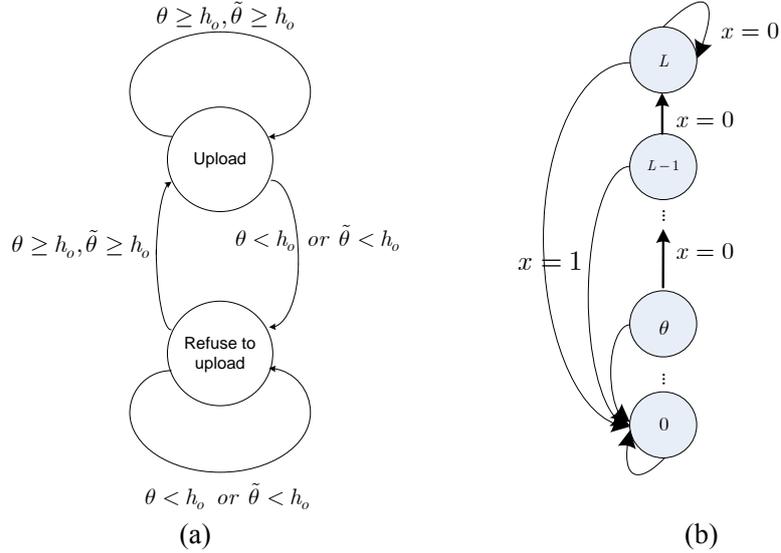

(a)      (b)

Figure 1. The schematic representation of a social norm

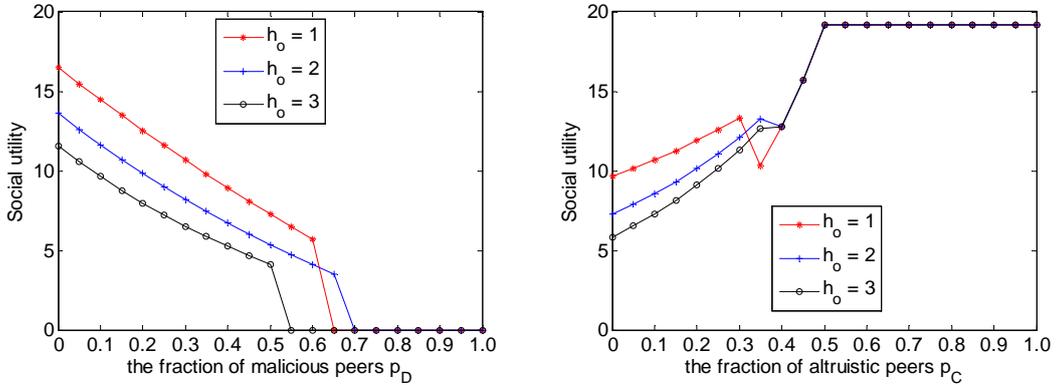

Figure 2. Average utility of reciprocative peers against $p_D$ and $p_C$

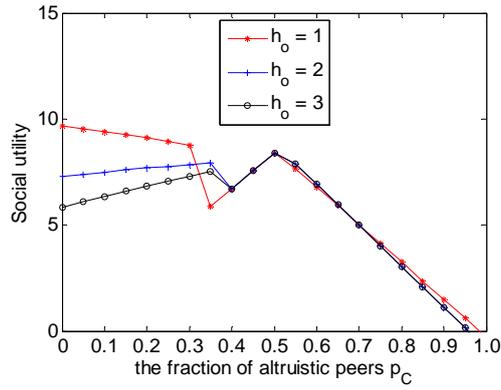

Figure 3.  Social utility of all peers against $p_C$

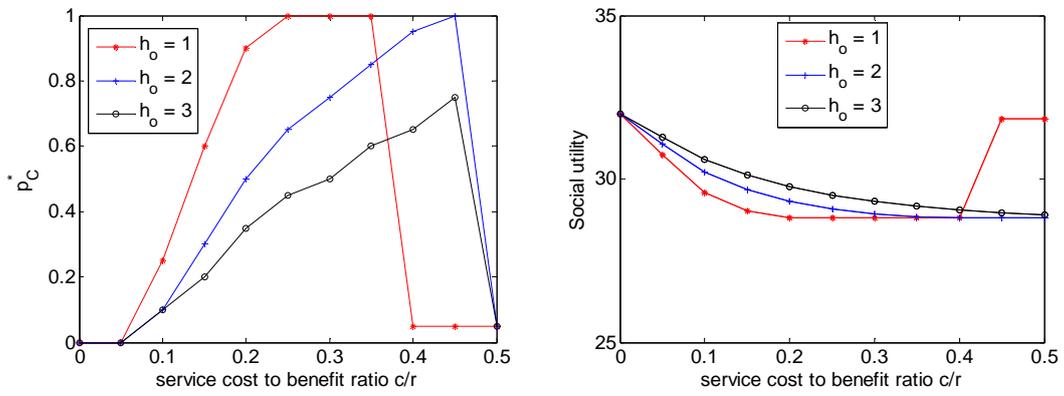

Figure 4.  The optimal fraction of altruistic peers $p_C^*$ and the corresponding social utility

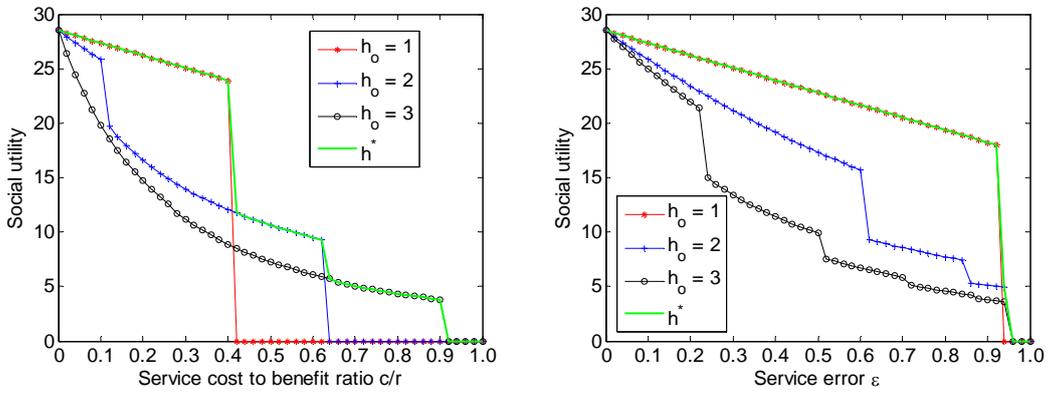

Figure 5.  The protocols' performance against $c/r$ and $\varepsilon$.

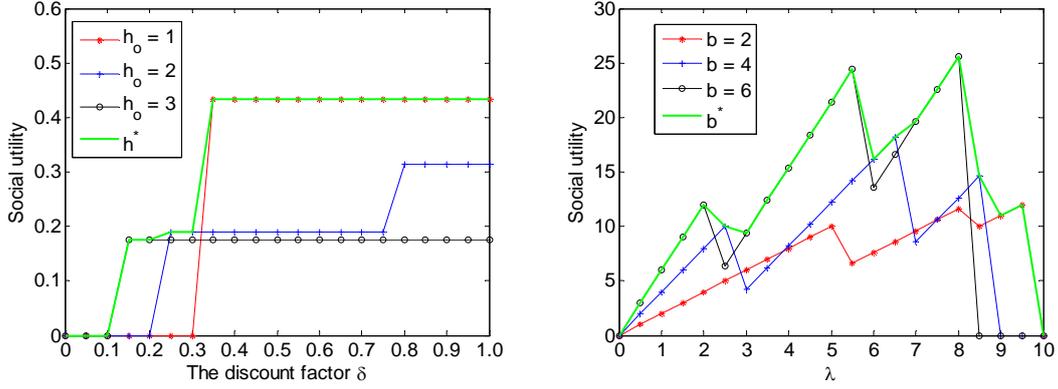

Figure 6. The protocols' performance against $\beta$ and $\lambda$

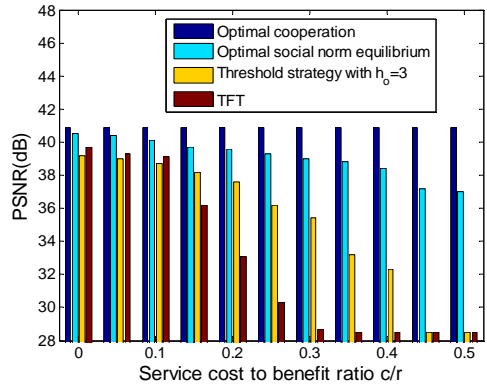

Figure 7. The PSNR of different protocols

APPENDIX F

TABLES

Table 1. The utility matrix of a gift-giving game

|  | Server | |
|---|---|---|
|  | S | NS |
| Client | r , c | 0, 0 |

Table 2. The sequence of events in one transaction

**Step 1:** A peer, denoted as $Peer_A$, sends a search request of a media chunk to the tracker.

**Step 2:** The tracker returns a list of peers who possess the chunk to $Peer_A$.

**Step 3:** $Peer_A$ randomly selects a peer $Peer_B$ from the list to send a download request.

**Step 4:** $Peer_B$ sends a request to the tracker to look up $Peer_A$'s reputation $\theta_A$.

**Step 5:** $Peer_B$ decides whether to upload the requested chunk to $Peer_A$.

 **If** ( $\theta_A \geq h_o$ & $Peer_B$ uploads the chunk) **or** ( $\theta_A < h_o$ & $Peer_B$ refuses to upload the chunk)

  $Peer_B$ is regarded as behaving well and $\phi$ outputs 0

 **Else**

  $\phi$ outputs 1

Table 3. The algorithm to solve the problem (OSNE)

1. **Input:** $(r,c,\varepsilon,\delta,\lambda,L)$ and $\hat{b}$, which is the maximum allowed value of $b$ in the system
2. If (18) is satisfied and $\delta > D$, the algorithm starts
3. **Initialize:** $h_o := 1$, $b := \hat{b}$, flag = 0
4. **Procedure:**
5. **While** ($h_o \leq L$) & (flag = 0) **do**
6.     **While** ($b \geq 1$) & (flag = 0) **do**
7.         Set flag = 1 if the protocol with $(h_o, b)$ is a social norm equilibrium
8.         $b := b - 1$
9.     **End While**
10.     $h_o := h_o + 1$
11. **End While**
12. **Set** $\left(h_o^{\#}, b^{\#}\right) := (h_o - 1, b + 1)$
13. **Find:** the optimal social norm equilibrium with $h_o^{\#}$, set as $\left(h_o^{\#}, \overline{b}\right)$
14. **Find:** the optimal social norm equilibrium with $b^{\#}$, set as $\left(\overline{h}_o, b^{\#}\right)$
15. **Return:** $\left(h_o^*, b^*\right) := \arg\max \left\{U_{\left(h_o^{\#}, b^{\#}\right)}, U_{\left(h_o^{\#}, \overline{b}\right)}, U_{\left(\overline{h}_o, b^{\#}\right)}\right\}$
16. **End Procedure**

Table 4. Decoded Video Quality

|  | Decoded Video Quality in PSNR (dB) | | | |
|---|---|---|---|---|
|  | $c/r = 0.1\ \varepsilon = 0.1$ | $c/r = 0.3\ \varepsilon = 0.1$ | $c/r = 0.1\ \varepsilon = 0.3$ | $c/r = 0.3\ \varepsilon = 0.3$ |
| $h_o = 1$ | 39.2 | 38.2 | 37.1 | 36.6 |
| $h_o = 2$ | 38.7 | 38.6 | 36.7 | 36.2 |
| $h_o = 3$ | 38.4 | 36.9 | 36.9 | 36.5 |
| TFT | 39.1 | 33.9 | 32.2 | 28.5 |

Table 5. The lookup table of the reputation scheme

| Current reputation $\theta$ | $y = 0$ | $y = 5$ | $y = 10$ | $y = 20$ |
|---|---|---|---|---|
| 0 | 0 | 1 | 2 | 3 |
| 1 | 0 | 2 | 3 | 3 |
| 2 | 0 | 3 | 3 | 3 |
| 3 | 0 | 3 | 3 | 3 |